**Insights into the Impact of COVID-19 on Bicycle Usage in Colorado Counties**

**Abdullah Kurkcu, Ph.D.**
Lead Traffic Modeler,
Ulteig,
5575 DTC Parkway, Suite 200, Greenwood Village, CO 80111, USA
E-mail: abdullah.kurkcu@ulteig.com

**Ilgin Gokasar, Ph.D.**
Associate Professor,
Department of Civil Engineering,
Bogazici University
Bebek-Istanbul, 34342
Email: ilgin.gokasar@boun.edu.tr

**Onur Kalan, Ph.D**
New York, NY, USA
Email: dronurkalan@gmail.com

**Alperen Timurogullari**
Department of Civil Engineering,
Bogazici University
Bebek-Istanbul, 34342
Email: timurogullarialperen@gmail.com

**Burak Altin**
Department of Civil Engineering,
Bogazici University
Bebek-Istanbul, 34342
Email: brkaltinn55@gmail.com

Word Count: 5457 words + 6 tables (250 words per table) = 6,957 words

Submitted on August 1st, 2020

## ABSTRACT

Coronavirus, which emerged in China towards the end of 2019 and subsequently influenced the whole world, has changed the daily lives of people to a great extent. In many parts of the world, in both cities and rural areas, people have been forced to stay home weeks. They have only been allowed to leave home for fundamental needs such as food and health needs, and most started to work from home. In this period, very few people, including essential workers, had to leave their homes. Avoiding social contact is proven to be the best method to reduce the spread of the novel Coronavirus. Because of the COVID-19 pandemic, people are adapting their behavior to this new reality, and it may change the type of public events people perform and how people go to these activities. Consumer behaviors have been altered during the pandemic. While people try to avoid gatherings, they also stayed away from mass transport modes and turned to private modes of transportation more -- private cars, private taxis and bike-sharing systems; even walking became more popular. In this study, we attempt to analyze how the use of bicycling has changed -- pre- and post-pandemic -- using open data sources and investigating how socio-economics characteristics affect this change. The results showed that average income, average education level, and total population are the most crucial variables for the Pandemic to Transition period and the Transition to the Normalization period.

**INTRODUCTION AND BACKGROUND**

In December 2019, the COVID-19 outbreak began in China and rapidly spread in numerous countries all over the world. The pandemic, which has infected millions of people and resulted in the death of thousands of people, was believed to originate from bats. Even though the origin has not yet been proven, the outbreak has contaminated nearly every country in the world and, at the time of this writing, is not been slowing down. It was swiftly declared a pandemic by the World Health Organization (WHO (*1*) in March 2020, proving the severity of the issue.

Many countries have taken extraordinary measures to prevent the spread of the virus as much as they reasonably could, shutting down public areas, restricting crowded events, and encouraging or imposing work from home orders. All of these measures fall under the umbrella of "social distancing." Social distancing is a dominant factor for preventing further spreading of the virus (such as COVID-19, SARS, MERS, etc.), transmitted by respiratory droplets, and requires some certain physical space between people (*2*). Some countries have taken stringent measures such as configuring lockdowns in specific regions of the country. In contrast, other states have enforced less strict standards (*3*). This also is true in the United States, where there is not a unified approach – where states are taking different approaches to containing the virus. Taking everything into account, one fact is clear that the effects of social distancing and coronavirus will last for a long time, and could become permanent.

Social distancing measures have affected consumer behavior, public gatherings, transportation modes, and, in general, have altered how people live their daily lives. For instance, many countries applied lockdowns, changing how people go to supermarkets, either limiting the capacity of an enclosed area or implementing various curfews to different age groups. Scheduled events involving many people were postponed; some production and factory processes were shut down. The gears of the economy, imports, and exports suddenly started to decrease swiftly, forming a feeling of anxiety for governments. Many people suddenly became unemployed, furloughed, or were forced to work from home, and most public activities were postponed.

As a result, travel demand suffered a dramatic drop, and many countries reported that their car traffic and the use of public transportation had drastically reduced (*4; 5*). The primary transportation mode between cities and countries – air flights – have been dramatically cut back as well, impacting the world's economy significantly. When social distancing measures are lifted, it is expected that out-of-home activities will regain their popularity and the use of public transportation will increase again.

However, as social distancing measures are lifted, some people will continue to be afraid of the infectiousness and fatality of the virus. At this point, people might prefer alternative travel modes that comply with social distancing guidelines (*6*). Certain features of transportation methodology will undergo augmentation, altering what we approach as norms to differ. The majority will prefer to ride a personal car, motorcycle, or bike in such cases. Because bikes are the most affordable option while providing the highest safety regarding social distancing measures due to less interaction with people, bike usage will be preferred by many people as an alternative to public transportation. This protects them from possible contamination and reassures an opinion of safety amongst people. Walking to nearby places is also a valid option; despite that, this might not apply safely to specific regions. For instance, walking in a more closed-in area in a crowded city or neighborhood still carries a great deal of risk of contamination, however in a more deserted community, walking should be the most efficient and safe way of travel.

In a research study performed by the Institute for Transportation and Development Policy (ITDP, 2020) in March in Guangzhou, China, 34% of people who used public transit systems continued to use public transportation, while 40% of people did not. They preferred to use private cars, taxis, walking and cycling because they were worried about the risk of getting infected. Another research study revealed that usage of bicycle-sharing systems in Beijing (China) increased by 150% from February 10th to March 4th, 2020 (*7*). Furthermore, short-distance riding increased as well as long-distance travel increased. According to some bike-sharing companies' data, the share of long-distance cycling usage (more than 3 km) increased nearly twice in comparison to ridership during the same period of last year (*8*). In England, some bike stores report facing increased demand. Manager of Broadribb Cycles says that sales increased from around 20 – 30 bikes a day to 50 bikes a day (*9*). Furthermore, cycling usage rate has seen a 200% increase, while car use roughly decreases by 40% in Manchester, England (*10*).

Recent studies conducted in various countries also investigated the change in bicycle usage during the COVID-19 pandemic. For example, a study about New York City's Citibike usage during the COVID-19 pandemic demonstrated that the bike-sharing system is much more resilient than urban transportation systems to disruptive events. The results indicated a 70% bike ridership drop versus 90% transit ridership drop, respectively. It also found that the average duration of Citibike travels significantly increased during the pandemic (*11*). Another study, a survey performed to inquire about behavioral changes in terms of bike usage in Sydney, Australia, revealed an increase in the willingness of bike usage for both hygiene and enjoyment reasons (*12*). In a Budapest, Hungary, case, modal share change during the COVID-19 pandemic period investigated. That study found that there are positive trends in the growing popularity of cycling as the share of public transport decreases (*13*). These recently published studies in the literature illustrate the importance of cycling and quantify the resilient nature of cycling as well as its preference over traditional public transit modes. Such a dramatic increase in cycling captured the attention of some state governments. For instance, Greater Manchester (England) announced that emergency funds will be allocated to promote walking and cycling to build a more resilient and stronger region, where the data shows that cycling increased 42% in Greater Manchester compared to the pre-lockdown period (*14; 15*).

Some researchers evaluated the increase in bicycle usage through a healthy community lens in the literature. Nyenhuis et al. (*16*) provide some insights about exercising during the pandemic. They quote an online forum user's indication about the Appalachian Trail as: "I have seen so many more people on bikes around town compared to last Spring. I am not just talking about road cyclists, but also people on cruiser bikes, mountain bikes, etc. It is actually good to see so many people out riding with their kids, spouses, etc." (*16*). Park, Kim and Lee (*17*) is another study, which emphasizes the health impacts of bicycle use during the COVID-19. Besides the physical health concerns, also the psychological health impacts of bike use during the COVID-19 is also inquired. Additionally, a difference in differences (DID) analysis is performed in Seoul, South Korea, for pre- and post-strict social distancing periods. The results show that average bike-sharing system usage among commuters and weekend users increased during social distancing period of 2020 (*17*). In an Oslo, Norway case, it is found that the recreational activities increased and cyclists appeared on trailways and in green areas much more than before the COVID-19 pandemic period (*18*).

As current trends and studies indicate, an increase in bicycle usage due to the COVID-19 pandemic might create an opportunity to increase sustainable transportation. Yet, an in-depth investigation is necessary to be able to understand the extent of this new normal so that the process can be controlled in favor of society. This study aims to reveal the relationship between bicycle usage and socioeconomic factors by comparing data collected before and after the COVID-19 pandemic. It uses continuously collected bicycle count data from fixed locations in the greater Denver metropolitan area before and after the pandemic to quantify the change in non-motorized traffic volumes. Therefore, bicycle catchment distances are taken as the radius by centering four bike count stations (*19*), and the socioeconomic data within these catchment areas are collected (*20*). Afterward, obtained socioeconomic data are analyzed with the change in bicycle usage percentages by using a machine learning tool, Partial Least Square Regression (PLSR), and initial results, and explanations are obtained in correlation with the COVID-19 pandemic.

## BICYCLE COUNT DATA

To better account for bicycle and pedestrian activity, the Colorado Department of Transportation (CDOT) implemented a statewide bicycle and pedestrian plan to expand counting locations to provide a comprehensive understanding of bicycle and pedestrian activity throughout the state in 2012. This plan identified appropriate uses and needs for non-motorized data, developed program goals, and created implementation plans. This study utilizes one of the outputs of the statewide plan -- continuous bicycle count data – openly available through CDOT's website. Figure 1 below illustrates the automated data collection methodology. This developed data pipeline is also publicly available, and the documentation of the code can be accessed at (*21*).

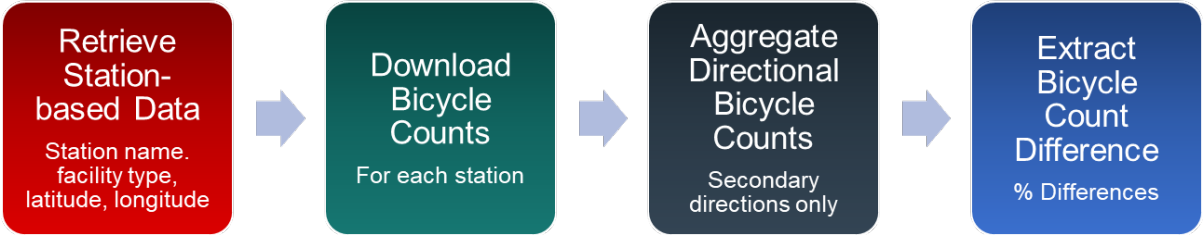

**Figure 1. Bicycle Count Data Collection Pipeline**

The data below in Table 1 is the ratio of change in bicycle counts, which was obtained using the data pipeline explained in Figure 1 for four bicycle stations in various Colorado counties, located in the USA. It consists of information for the percentage change in bicycle usage by dividing the periods as Pre-Pandemic, Pandemic, Transition, and Normalization. The location of each station is used as the center, and circles are drawn around stations to define the areas affected by the stations. The radius of the circles is determined by the APTA standards and known as a catchment distance (*19*). Figure 2 shows bicycle count stations and their catchment areas with various selected buffer sizes. The socio-demographic data, which consists of income, education, age, population, and gender information, was collected for each catchment area corresponding to a station.

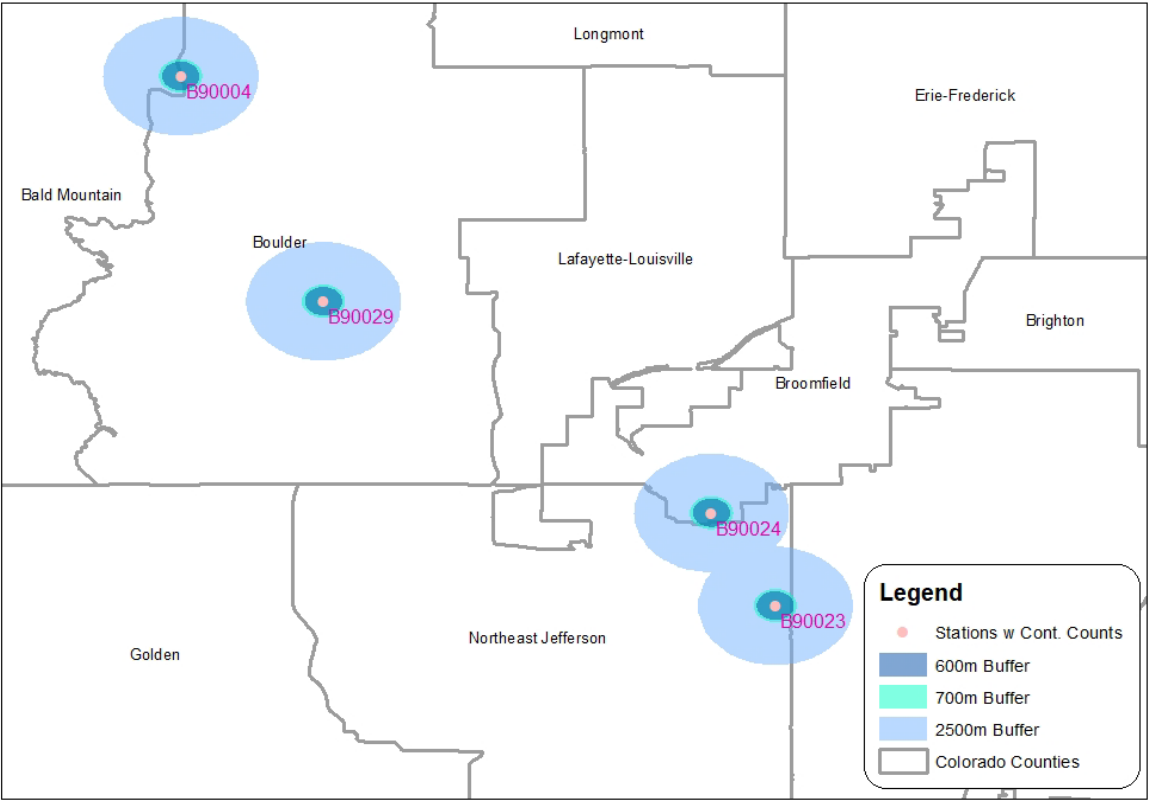

**Figure 2. Selected Bicycle Stations and Catchment Areas**

After circumscribing the catchment area of each station, the counties within these regions are identified, and the required data for these counties are collected from the database of Census Bureau of U.S. Raw data containing economics, educational, and demographics data of counties' households. These data are categorized, and average income level, average educational level, and average age are obtained.

- Economics data is divided into ten different categories and the number of people multiplied by these income categories and divided by the total number of households to achieve an average income level for each county.
- Educational data is divided into nine different groups, and these categories are numbered 0 to 8, respectively, and the number of households multiplied by these numbers and distributed by the total number of households to get average education level.
- The same procedure is followed for demographics data to obtain average age for each county. Furthermore, total population and male-to-female ratio are taken from raw data. After that, using these data, average income level, educational level, age, total population, and male to female ratio are obtained for each station.

**Table 1. Data used in the analysis**

| Station | Pre-Pandemic to Pandemic (%) | Pandemic to Transition (%) | Transition to Normalization (%) | Avg. Income | Avg. Education | Avg. Age | Total Population | Male / Female |
|---|---|---|---|---|---|---|---|---|
| **0** | 0.36 | 1.70 | -1.28 | 0.65 | 0.84 | -0.41 | -0.81 | 1.71 |
| **1** | 0.87 | 1.80 | -1.55 | 0.79 | 0.96 | -0.56 | -0.91 | -0.36 |
| **2** | -1.70 | -0.55 | -0.90 | 0.27 | -0.29 | 1.72 | 0.14 | -0.84 |
| **3** | 0.47 | -0.84 | 3.73 | -1.70 | -1.51 | -0.75 | 1.58 | -0.50 |

**METHODOLOGY**

In this study, using the change in the two-year bicycle trip rate for the previous period, the two-year change rate of bicycle trips of the following month will be estimated. The flow chart showing the data collection, preprocessing, and the prediction model is given in Figure 3.

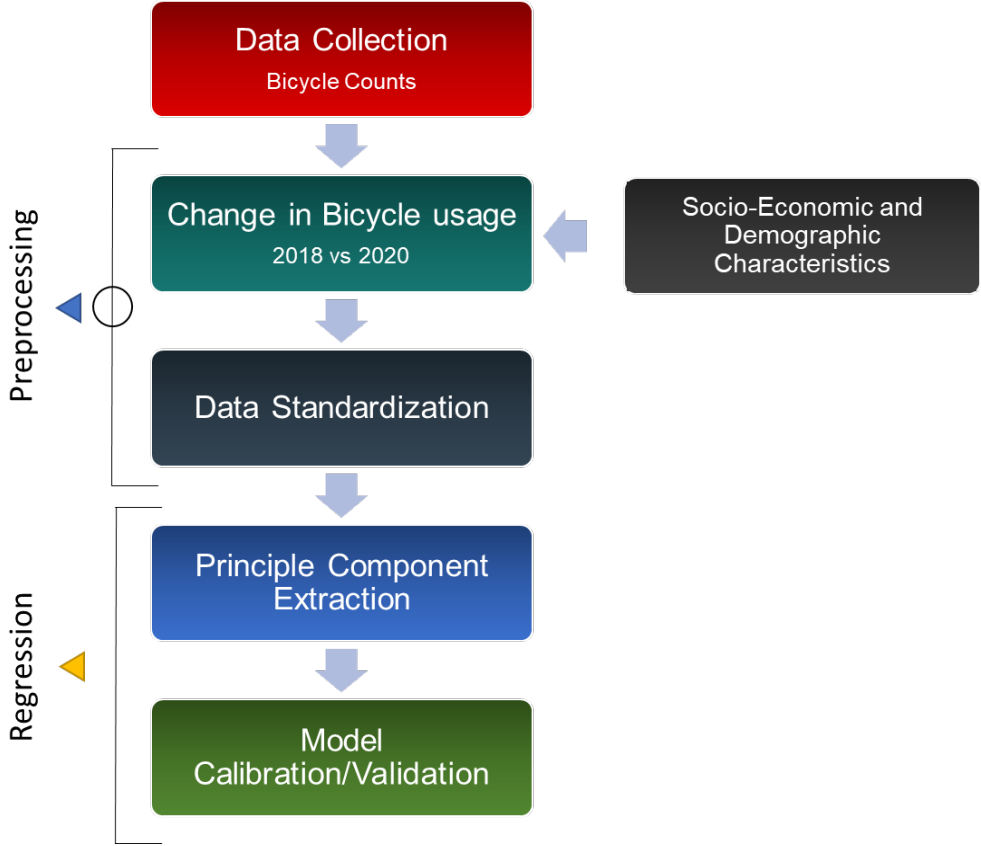

**Figure 3. Data Analysis Flowchart**

The available data includes the age, income level, education level, gender, and population information of the users in a station's coverage area, and the change rate of the number of bicycle trips in two years per period. The rate of change was obtained by dividing the number of bicycle trips at each station in 2020 by that in 2018. The calculation of the periodic change rate, which is the dependent variable of this study, is shown in **Equation 1**.

$$r_{t,t-1} = \frac{n_t}{n_{t-1}} \qquad (1)$$

where:

1. $r_{t,t-1}$ is the change rate of bicycle usage at a station from period t-1 to period t.
2. $n_t$ is the number of bicycle usage at the station in period t
3. $n_{t-1}$ is the number of bicycle usage at the station in period t-1

The data on bicycle trip rates and demographic characteristics of four different stations includes each transition phase between periods. Since there is a small number of data, it is determined to apply the Partial Least Square Regression (PLSR) method. PLSR is a suitable method for a small amount of data analysis, to observe the change in periodic change rate according to demographic characteristics.

To process the data appropriately in the PLSR method, the original data ($k$) must be transformed into standardized data ($x$). Standardization for each variable is performed as follows:

$$x_{i,j} = \frac{k_{i,j} - \bar{k}_j}{\sigma_{k,j}} \qquad (2)$$

where:

1. $x_{i,j}$ is the standardized one-dimensional data point at row $i$ and column $j$ of $x$,

2. $k_{i,j}$ is the one-dimensional data in the k at row $i$ and column $j$,

3. $\bar{k}_j$ is the mean value of the column $j$ of the original data

4. $\sigma_{k,j}$ is the standard deviation of column $j$ of $k$

After the dependent variable calculation and data standardization is completed, the data is ready to be used in PLSR analysis. PLSR is a machine-learning technique that integrates assumptions and attributes of Principal Component Analysis (PCA) and multiple linear regression. In the PLSR method, X (the data set of independent variables) and Y (the data set of dependent variables) sets are written as a matrix product of factors, which are perpendicular to each other and variable-specific loads in a different space. The Nonlinear Iterative Partial Least Squares (NIPALS) algorithm is selected to apply decomposition to the data. The decomposition procedure is shown below

$$X = T \cdot P^T + E \tag{3}$$

$$Y = U \cdot Q^T + F \tag{4}$$

where:

$$T = X \cdot W_x \tag{5}$$

$$U = Y \cdot W_y \tag{6}$$

where:

- $X$ and Y are the set of independent variables and dependent variables, respectively,
- $T$ and $U$ are the component matrix of $X$ and $Y$, respectively. The shape of component matrix changes according to the selected number of components.
- $P$ and $Q$ are the loading matrix of $X$ and $Y$, respectively,
- $E$ and $F$ are the residual errors of $X$ and $Y$, respectively,
- $W_x$ and $W_y$ are the weight matrices of $X$ and $Y$, respectively.

Following decomposition of X and Y is completed, iteration is performed until T and U matrices are obtained, which will maximize the covariance's of X and Y. The first step of the iteration is equalizing the errors to X and Y such that E = X and F = Y. The algorithm of minimizing error terms for one variable is given below:

```
t = e_j
u = f_j
while True:
    p = (X^T u) / ||X^T u||
    t = Xp
    q = (Y^T t) / ||Y^T t||
    u = Yq
    if t does not change:
        break
E = E - tp^T
F = F - uq^T
```

where:

- $e_j$ and $f_j$ are some column of E and F, respectively
- t and u are the principal components of T and U, respectively
- p and q are the principal components of P and Q, respectively.

The same steps are repeated for each variable of X and Y. In this study, the shape of X is 4x5, and the shape of Y is 4x1. After the iterations are finished, factor matrices (T and U) and weight matrices (P and Q) are created established to maximize the covariance between T and U.

**RESULTS**

Before interpreting the outputs resulting from the PLS regression analysis, the relationship between each variable and the dependent variable was visually examined. Figure 4 shows the relationship between average income and change in the number of bicycle trips for each period. The average income shows a negative correlation with the periodic change rate for pandemic and normalization periods, where the relationship is in contrast in the transition period. This means that as the income level of people increases, the bicycle usage rate decreases. As shown before, the reason for this can be that the possibility of having a private car is higher for people with higher incomes (*22*). Besides, as the income level increase, the preferences might change in favor of private on-demand services like taxis.

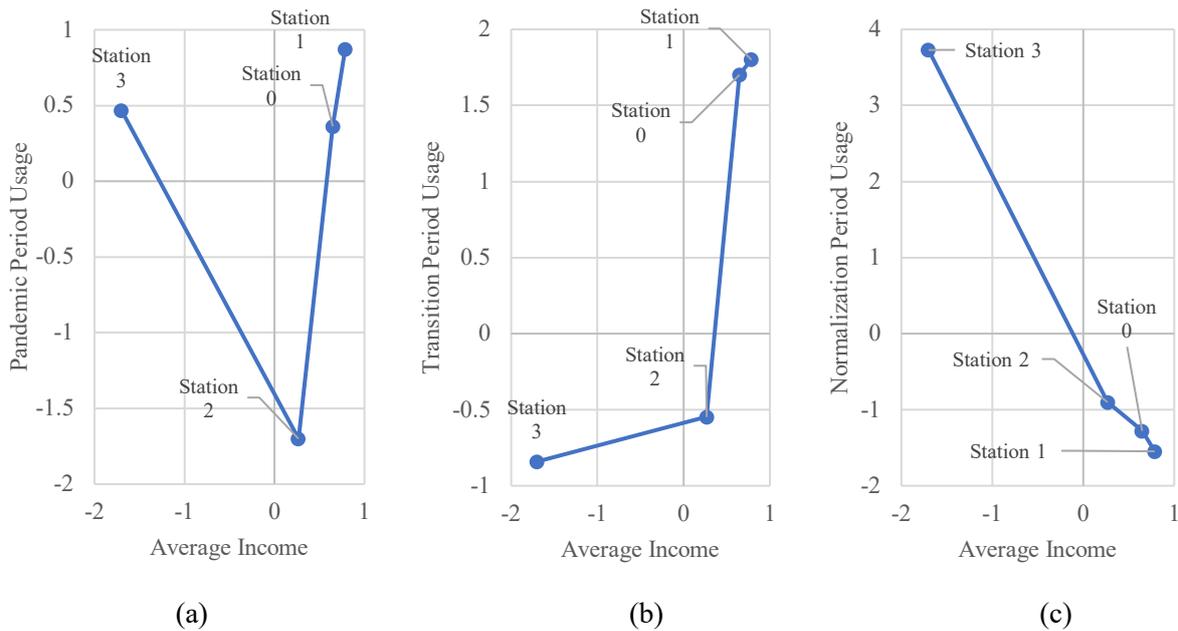

|        (a)        |        (b)        |        (c)        |

**Figure 4. "Average income" / "Periodic change rate" plots for Pre-pandemic to Pandemic (a), Pandemic to Transition (b) and Transition to Normalization (c) phases**

Figure 5 demonstrates that the average education level has a negative correlation with the bicycle usage rate in the normalization period. According to a study, as average education level rises, the welfare level of people also increases (*23*). So that the likelihood of having private cars increases, and this removes the necessity to use other transportation systems such as public transportation, bikes, or walking. Even though the education level has a negative correlation during Normalization, the situation is the opposite for the transition period that is the bicycle usage rate in the transition period increases in parallel with the education level. Meaning that educated people are more sensitive about the pandemic, and they avoid crowded places where the risk of the rate of contamination is higher

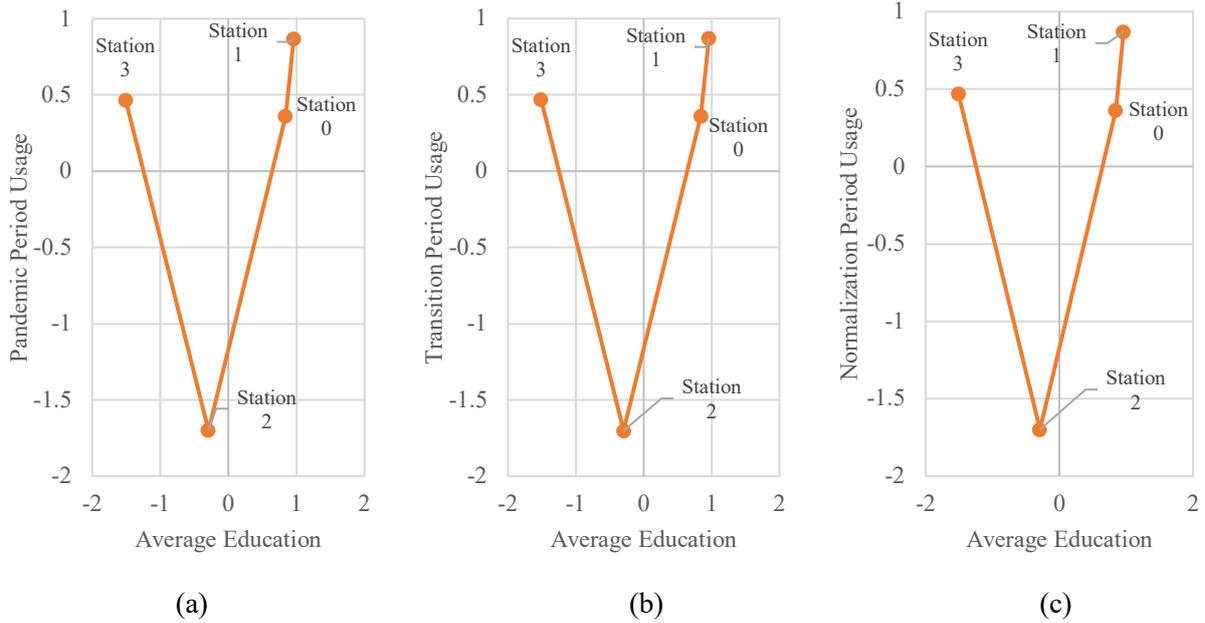

(a)                 (b)                 (c)

**Figure 5. "Average education level" / "Periodic change rate" plots for Pre-pandemic to Pandemic (a), Pandemic to Transition (b) and Transition to Normalization (c) phases**

Figure 6 shows that there is a general decrease in bicycle usage as the age increases except for the normalization period. The decrease is especially significant during the pandemic period. It might have different reasoning, such as the decrease in physical ability, less sensitivity, and conciseness for the situation or not using any means of transportation due to infection risk.

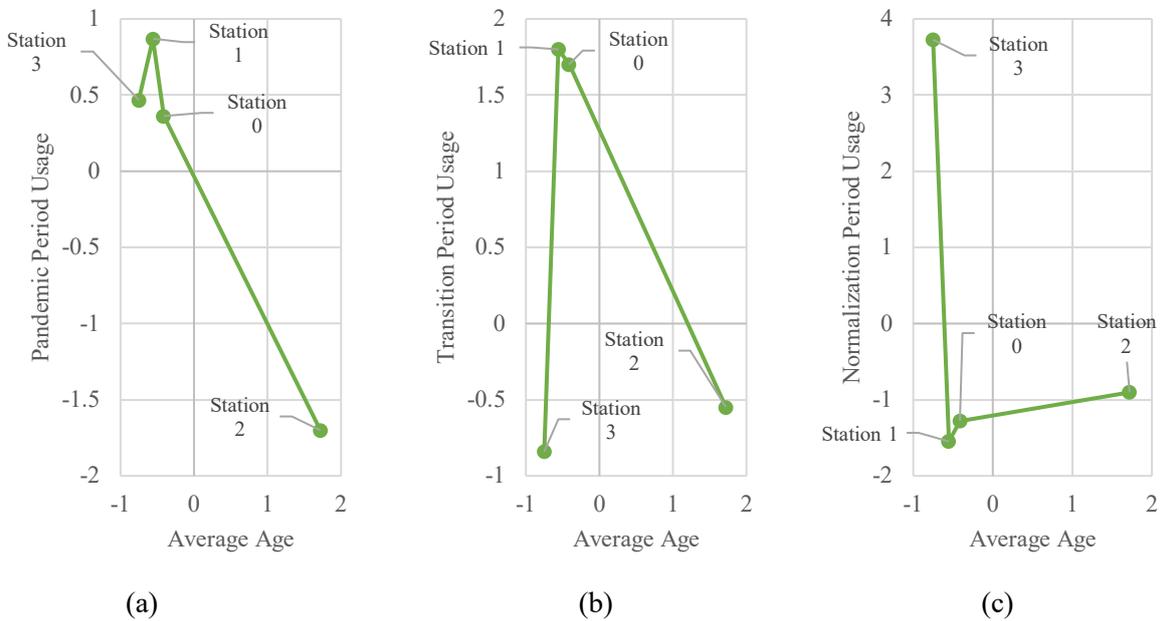

(a)                 (b)                 (c)

**Figure 6. "Average age" / "Periodic change rate" plots for Pre-pandemic to Pandemic (a), Pandemic to Transition (b) and Transition to normalization (c) phases**

According to Figure 7, the total population has a positive correlation with the bicycle usage rate in the normalization period, meaning that as the overall population increases, bicycle usage rate increases. So, people stay away from crowded transportation options where the risk is higher for higher population areas to avoid getting infected. Thus, the tendency to use a bicycle as transportation options increases.

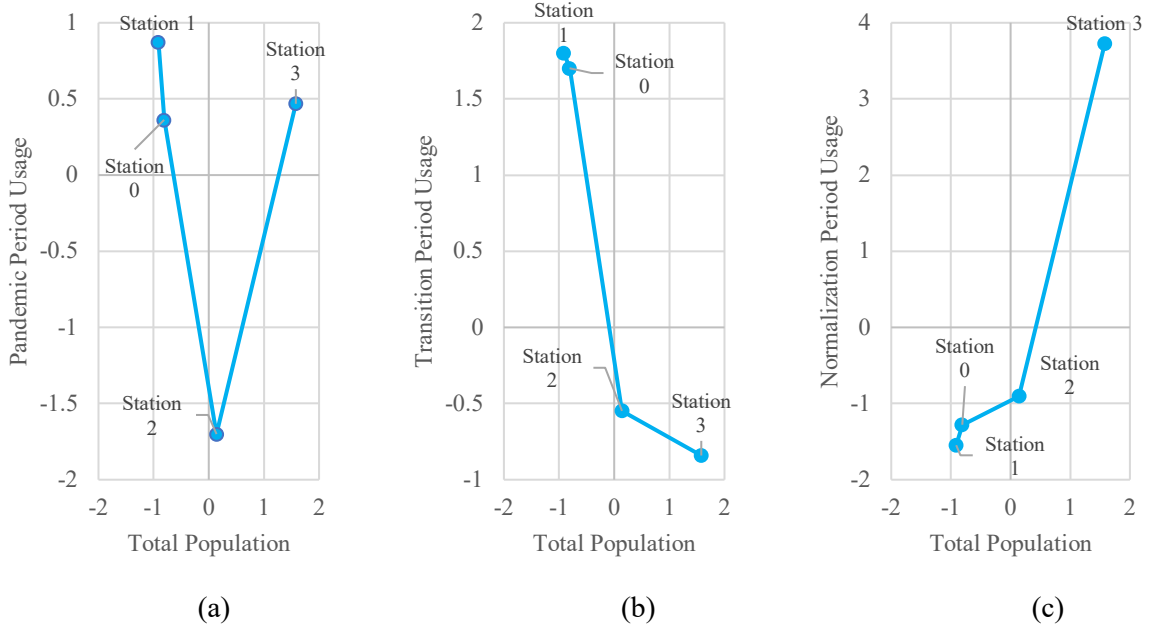

(a)            (b)            (c)

**Figure 7. "Total population" / "Periodic change rate" plots for Pre-pandemic to pandemic (a), Pandemic to Transition (b) and Transition to normalization (c) phases**

Figure 8 demonstrates that there is no correlation between change in bicycle usage and the male/female rate of the county.

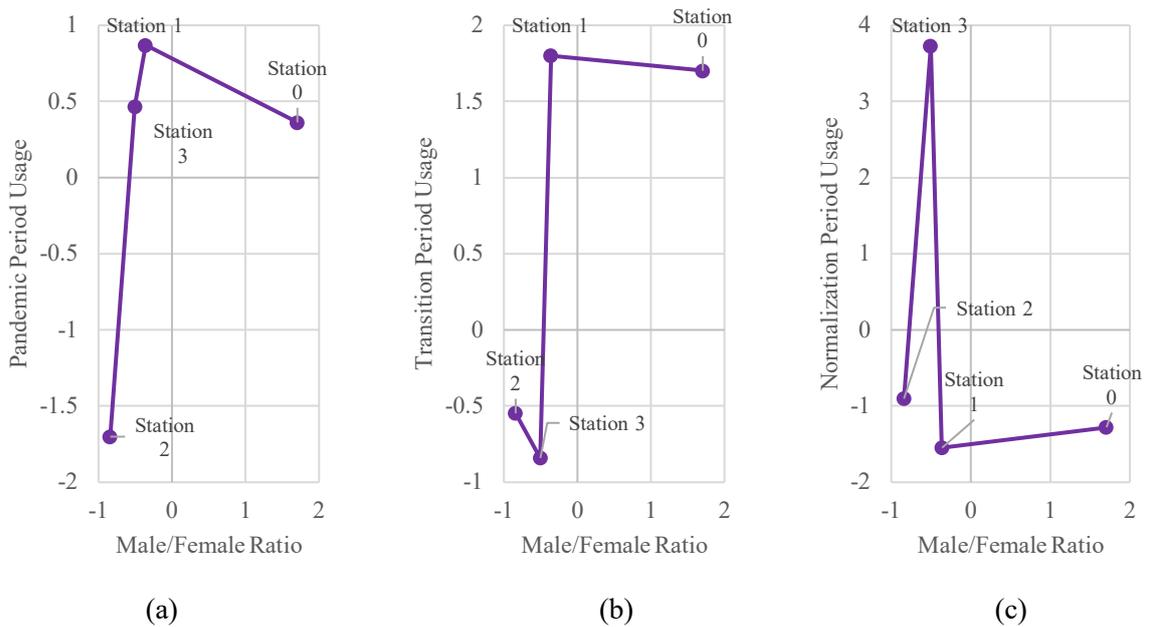

(a)            (b)            (c)

**Figure 8. "Male/female rate" / "Periodic change rate" plots for Pre-pandemic to Pandemic (a), Pandemic to Transition (b) and Transition to Normalization (c) phases**

**Proportion of Variance**

Table 2 summarizes the proportion of variance explained matrices of the PLSR for Pre-Pandemic-to-Pandemic, Pandemic-to-Transition, and Transition-to- Normalization period, respectively. For all three periods, the 3rd latent factor has the least X variance value. This latent factor can also make the prediction model to overfit. A well-fitting model would have similar and high X and Y variance values. Additionally, as the Y variance value increases, for a new sample of dependent values, the latent factor becomes more powerful in terms of explaining the variation. The more a latent factor can explain the variance of independent variables, the observed values of the set of independent variables are reflected even better. Latent factor 1, for all three periods, have the highest X variance and Y variance and, therefore, is more powerful for the developed model. According to the adjusted R-square values, the Transition-to-Normalization period is the best-fitted period for this model, and the Pre-Pandemic-to-Pandemic period is the worst-fitted period for this model.

**Table 2. Proportion of Variance Explained Matrices of the PLSR for Pre-pandemic to Pandemic periods**

|  | Statistics | Latent Factors | | |
|---|---|---|---|---|
|  |  | 1 | 2 | 3 |
| *Pre Pandemic to Pandemic* | X Variance | 0.330 | 0.555 | 0.115 |
|  | Cumulative X Variance | 0.330 | 0.885 | 1.000 |
|  | Y Variance | 0.836 | 0.088 | 0.076 |
|  | Cumulative Y Variance | 0.836 | 0.924 | 1.000 |
|  | Adjust R- Square | 0.754 | 0.773 | 0.000 |
| *Pandemic to Transition* | X Variance | 0.637 | 0.260 | 0.104 |
|  | Cumulative X Variance | 0.637 | 0.896 | 1.000 |
|  | Y Variance | 0.921 | 0.061 | 0.018 |
|  | Cumulative Y Variance | 0.921 | 0.982 | 1.000 |
|  | Adjust R- Square | 0.882 | 0.947 | 0.000 |
| *Transition to Normalization* | X Variance | 0.635 | 0.280 | 0.085 |
|  | Cumulative X Variance | 0.635 | 0.915 | 1.000 |
|  | Y Variance | 0.950 | 0.050 | 9.54E-02 |
|  | Cumulative Y Variance | 0.950 | 1.000 | 1.000 |
|  | Adjust R- Square | 0.925 | 1.000 | 0.000 |

**Weights and Loadings:**

5x3 $W\_x$ and 1x3 $W\_y$ weight matrices for all period transitions are given in Table 3. These matrices are positively correlated with the T and U matrices in line with the (3) and (4). The X-weights in Table 3 demonstrate the correlation of X variables with the Y scores. 5x3 P and 1x3 Q loading matrices are also shown in Table 3. The X-loadings in Table 4 also represents the direction of the principal components for each independent variable in X-space. According to the preliminary assessment of the loadings and weights tables, the results we expect from the interpretation of variance coefficients table are:

- Average age and gender are the dominating demographical parameter for the Pre-Pandemic-to-Pandemic period. An increase in age and the male rate in population will cause a decrease in the change of bicycle trips.
- In contrast to the Pre-Pandemic-to-Pandemic period, in the Pandemic-to-Transition period, the male/female rate shows a strong positive correlation with the change of bicycle trips on stations. From the loading matrix, it is also expected that there is a positive correlation with average age and the dependent variable.
- At this period, it is expected that there is a strong negative correlation between average income, education level, and age and the periodic change rate. In contrast, the total population and male/female rate shows a positive correlation.

**Table 3. Weight Matrices of the PLSR for all period transitions**

| Variables | Latent Factors | | | | | | | | |
|---|---|---|---|---|---|---|---|---|---|
|  | 1 | 2 | 3 | 1 | 2 | 3 | 1 | 2 | 3 |
| Average Income | -0.077 | -0.477 | 0.168 | 0.457 | -0.241 | 0.135 | -0.580 | -0.212 | -0.073 |
| Average Education Level | 0.215 | -0.275 | 0.290 | 0.553 | 0.063 | 0.239 | -0.533 | 0.051 | -0.225 |
| Average Age | -0.900 | -0.721 | -0.475 | -0.242 | -0.980 | -0.420 | -0.199 | -0.789 | 0.561 |
| Total Population | -0.135 | 0.341 | -0.246 | -0.533 | 0.025 | -0.200 | 0.552 | 0.019 | 0.174 |
| Male/Female Rate | 0.345 | -0.482 | -0.924 | 0.377 | 0.022 | -0.917 | -0.185 | 0.594 | 0.775 |
| Dependent Variable Weight | 0.770 | 0.197 | 0.362 | 0.545 | 0.231 | 0.184 | 0.554 | 0.188 | 0.015 |
| Period | Pre-pandemic to Pandemic | | | Pandemic to Transition | | | Transition to Normalization | | |

**Table 4. Loading Matrices of the PLSR for all period transitions**

| Variables | Latent Factors | | | | | | | | |
|---|---|---|---|---|---|---|---|---|---|
| | 1 | 2 | 3 | 1 | 2 | 3 | 1 | 2 | 3 |
| Average Income | 0.108 | -0.627 | 0.390 | 0.506 | -0.393 | 0.222 | -0.562 | -0.121 | -0.068 |
| Average Education Level | 0.366 | -0.558 | 0.418 | 0.557 | -0.102 | 0.217 | -0.554 | 0.139 | -0.226 |
| Average Age | -0.756 | -0.283 | -0.140 | -0.095 | -0.918 | -0.067 | -0.083 | -0.773 | 0.580 |
| Total Population | -0.299 | 0.584 | -0.404 | -0.550 | 0.184 | -0.209 | 0.562 | -0.070 | 0.173 |
| Male/Female Rate | 0.603 | -0.300 | -0.700 | 0.383 | 0.296 | -0.925 | -0.280 | 0.604 | 0.760 |
| Dependent Variable Loading | 1.000 | 1.000 | 1.000 | 1.000 | 1.000 | 1.000 | 1.000 | 1.000 | 1.000 |
| Period | Pre-pandemic to Pandemic | | | Pandemic to Transition | | | Transition to Normalization | | |

**Parameters and Variable Importance**

Table 5 shows the importance of the independent variables for projection for its latent factor and period, and in Table 6, the regression parameters' estimated coefficients are presented for all periods. According to Table 5, in the pre-pandemic to pandemic period, the average age variable is the most critical variable for all latent factors. It has a much higher impact on the rate of bicycle use. From Table 6, it can be observed that the average age parameter has a negative effect on the dependent variable. Average income, average education level, and total population are the most crucial variables for the Pandemic-to-Transition period and the Transition-to-Normalization period. Average income and average education level parameters have a positive influence on the dependent variable, whereas the total population variable has a negative impact.

**Table 5. Variable Importance Matrices of the PLSR for all period transitions**

| Variables | Latent Factors | | | | | | | | |
|---|---|---|---|---|---|---|---|---|---|
| | 1 | 2 | 3 | 1 | 2 | 3 | 1 | 2 | 3 |
| Average Income | 0.171 | 0.368 | 0.369 | 1.023 | 1.000 | 0.992 | 1.298 | 1.269 | 1.269 |
| Average Education Level | 0.482 | 0.496 | 0.509 | 1.237 | 1.198 | 1.190 | 1.193 | 1.163 | 1.163 |
| Average Age | 2.013 | 1.978 | 1.924 | 0.542 | 0.758 | 0.762 | 0.446 | 0.586 | 0.587 |
| Total Population | 0.303 | 0.372 | 0.388 | 1.191 | 1.153 | 1.144 | 1.235 | 1.204 | 1.203 |
| Male/Female Rate | 0.771 | 0.805 | 0.960 | 0.844 | 0.817 | 0.855 | 0.413 | 0.499 | 0.500 |
| Period | Pre-pandemic to Pandemic | | | Pandemic to Transition | | | Transition to Normalization | | |

**Table 6. Parameters Matrices of the PLSR for all period transitions**

| Variables | Dependent Variable | Dependent Variable | Dependent Variable |
|---|---|---|---|
| Constant | 1.857E-15 | 0.528 | 4.286E-16 |
| Average Income | -0.092 | 0.268 | -0.784 |
| Average Education Level | 0.217 | 0.442 | -0.626 |
| Average Age | -1.007 | -0.535 | -0.543 |
| Total Population | -0.127 | -0.394 | 0.675 |
| Male/Female Rate | -0.164 | 0.051 | 0.046 |
| Period | Pre-pandemic to Pandemic | Pandemic to Transition | Transition to Normalization |

## CONCLUSIONS

Various studies in the literature suggest many people are turning to cycling as a resilient, safe, and reliable mode of transportation to fill the gap of safe transport modes. Many urban cycling networks have seen a dramatic increase in non-motorized traffic, including China, Hungary, Korea, and the United States.

Some governments are responding to this emerging demand by opening emergency bicycle lanes, converting roads, and opening them for the use of non-motorized traffic, and creating one-way bike lanes in order to comply with social distancing guidelines.

This study attempts to quantify such trends of increasing demand and investigate the relationship between bicycle usage and socioeconomic factors of surrounding areas. Three different study periods were created to understand the shift from pre-pandemic to post-pandemic levels. A machine learning tool, PLSR, was deployed to explain the underlying contributing factors to bike-demand. Initial results showed that average income, average education level, and total population are the most crucial variables for the Pandemic-to- Transition period and the Transition-to-Normalization period. Average income and average education level parameters have a positive influence on the dependent variable, whereas the total population variable has a negative impact.

This analysis has also shown the location-based impact of the COVID-19 pandemic on bicycle demand. While some stations experience an increased demand during the selected periods, others saw a decrease in bicycle usage.

The results of this study may be considered as evidence on the role of non-motorized travel modes -- cycling specifically. Even though results indicate that post-pandemic bicycle demand is slightly decreasing, an increasing number of cases may have an impact on future bicycle usage. Future work will focus on collecting additional data to cover the whole period of the spread of coronavirus in not only Colorado but also additional locations throughout the country. It will investigate the location-based impact on commuting needs and the long-term impacts of the virus as opposed to the limited duration used in this study.

## ACKNOWLEDGMENTS

The work in this paper is sponsored by Ulteig. The contents of this paper reflect the views of the authors who are responsible for the facts and do not represent any official views of any sponsoring organizations or agencies. The authors would also like to thank the Colorado Department of Transportation (CDOT) for making bicycle count data available.

**AUTHOR CONTRIBUTIONS**

The authors confirm contribution to the paper as follows: study conception and design: Ilgin Gokasar; data collection: Abdullah Kurkcu, Ilgin Gokasar, Alperen Timurogullari, Burak Altin; analysis and interpretation of results: , Ilgin Gokasar, Alperen Timurogullari, Burak Altin, Abdullah Kurkcu; draft manuscript preparation: Ilgin Gokasar, Alperen Timurogullari, Burak Altin, Abdullah Kurkcu, and Onur Kalan. All authors reviewed the results and approved the final version of the manuscript.